 \documentclass[12pt,preprint]{aastex}

 \usepackage{emulateapj5}

\slugcomment{to appear in the Astrophysical Journal}

\shorttitle{Light Curve of V1974 Cygni}
\shortauthors{Hachisu \& Kato}

\begin{document}

\title{Toward a unified light curve model for multi-wavelength 
observations of V1974 Cygni (Nova Cygni 1992)}


\author{Izumi Hachisu}
\affil{Department of Earth Science and Astronomy, 
College of Arts and Sciences, University of Tokyo,
Komaba, Meguro-ku, Tokyo 153-8902, Japan} 
\email{hachisu@chianti.c.u-tokyo.ac.jp}

\and

\author{Mariko Kato}
\affil{Department of Astronomy, Keio University, 
Hiyoshi, Kouhoku-ku, Yokohama 223-8521, Japan} 
\email{mariko@educ.cc.keio.ac.jp}

%
%



\begin{abstract}
     We present a unified model for optical, ultraviolet (UV),
and X-ray light curves of V1974 Cygni (Nova Cygni 1992).
Based on an optically thick wind model of nova outbursts,
we have calculated light curves and searched for the best fit model
that is consistent with optical, UV, and X-ray observations.
Our best fit model is a white dwarf (WD) of mass $1.05~M_\sun$
with a chemical composition of $X=0.46$, $C+N+O=0.15$, 
and $Ne = 0.05$ by mass weight.
Both supersoft X-ray and continuum UV 1455 \AA ~light curves
are well reproduced.  Supersoft X-rays emerged on day $\sim 250$
after outburst, which is naturally explained by our model: our
optically thick winds cease on day 245 and supersoft X-rays emerge
from self-absorption by the winds.  The X-ray flux keeps a constant
peak value for $\sim 300$ days followed by a quick decay on day 
$\sim 600$.  The duration of X-ray flat peak is well reproduced
by a steady hydrogen shell burning on the WD.
Optical light curve is also explained by the same model if we 
introduce free-free emission from optically thin ejecta.
A $t^{-1.5}$ slope of the observed optical and infrared fluxes 
is very close to the slope of our modeled free-free light curve
during the optically thick wind phase.  Once the wind stops, 
optical and infrared fluxes should follow a $t^{-3}$ slope,
derived from a constant mass of expanding ejecta.
An abrupt transition from a $t^{-1.5}$ slope to a $t^{-3}$ slope 
at day $\sim 200$ is naturally explained by the change from 
the wind phase to the post-wind phase on day $\sim 200$.
The development of hard X-ray flux is also reasonably 
understood as shock-origin between the wind and the companion star.
The distance to V1974 Cyg is estimated to be $\sim 1.7$~kpc with
$E(B-V)= 0.32$ from the light curve fitting for the continuum 
UV 1455 \AA~.
\end{abstract}




\section{Introduction}
     It has been widely accepted that 
classical novae are a thermonuclear runaway event 
on a mass-accreting white dwarf (WD).
Characteristic properties on nova evolution have been
understood from its ignition through the end of nuclear burning
\citep[e.g.,][for a review]{war95}.
The next step we need is quantitative studies of individual objects.
For instance, fitting of multi-wavelength light curves with
theoretical models enables us to determine nova parameters.
Such a work has been developed in the recurrent novae 
\citep[e.g.,][]{hac01ka, hac01kb, hkkm00, hac03a},
but not yet in the classical novae except
a pioneering work on V1668 Cyg (Nova Cygni 1978) by \citet{kat94}.  

V1974 Cygni (Nova Cygni 1992) is a best example for such studies
because it was extensively observed in all the wavelengths from 
$\gamma$-ray to radio.  
Among various observational data, three bands of optical, 
continuum ultraviolet (UV) at 1455 \AA~, 
and X-ray are used for our present study.
Based on an optically thick wind model of nova 
outbursts \citep[e.g.,][]{kat94h},
we try to develop a unified model that yields light curves
for each wavelength band.  The next section introduces the light curve
analysis based on our optically thick wind model.  In \S 3, 
we describe light curve fittings with X-ray, UV, and optical bands.
Discussion follows in \S 4.

\section{Modeling of V1974 Cyg}

\subsection{Optically thick wind model}
     After a thermonuclear runaway sets in on a mass-accreting WD,
its envelope expands greatly to $R_{\rm ph} \gtrsim 100 ~R_\sun$
and settles in a steady-state.  The decay phase of nova can be 
followed by a sequence of steady state solutions \citep[e.g.,][]{kat94h}.  
Using the same method and numerical techniques as in \citet{kat94h},
we have calculated theoretical light curves.

     We solve a set of equations, i.e., the continuity, equation
of motion, radiative diffusion, and conservation of energy, 
from the bottom of the hydrogen-rich envelope through the photosphere,
under the condition that the solution goes through a critical point
of steady-state winds.  The winds are accelerated deep inside 
the photosphere so that they are called ``optically thick winds.''
We have used updated OPAL opacities \citep{igl96}.  
We simply assume that photons are emitted at the photosphere
as a blackbody with the photospheric temperature of $T_{\rm ph}$.
Physical properties of these wind solutions have already been
published \citep[e.g.,][]{hac01ka, hac01kb, hac04k, hkn96,
hkn99, hknu99, hkkm00, hac03a, kat83, kat97, kat99}.
It should be noticed that a large number of meshes, i.e., 
more than several thousands grids, are adopted for the wind solutions
in an expanded stage of $R_{\rm ph} \sim 100 ~R_\sun$.

     Optically thick winds stop after a large part of the envelope
is blown in the winds.  The envelope settles into a hydrostatic
equilibrium where its mass is decreasing in time by nuclear burning.
Then we solve equation of static balance instead of equation of motion.
When the nuclear burning decays, the WD enters a cooling phase, in
which the luminosity is supplied with heat flow from the ash of 
hydrogen burning.

\subsection{Multiwavelength light curves}
     In the optically thick wind model, a large part of 
the envelope is ejected continuously for a relatively 
long period \citep[e.g.,][]{kat94h}.  After the maximum expansion
of the photosphere, its photospheric radius gradually decreases
keeping the total luminosity ($L_{\rm ph}$) almost constant.
The photospheric temperature ($T_{\rm ph}$) increases in time
because of $L_{\rm ph} = 4 \pi R_{\rm ph}^2 \sigma T_{\rm ph}^4$.
The main emitting wavelength of radiation moves from optical
to supersoft X-ray through UV.
This causes the decrease in optical luminosity 
and the increase in UV.  Then the UV flux reaches a maximum.
Finally the supersoft X-ray flux increases after the UV flux decays.
These timescales depend on WD parameters such as the WD mass and
chemical composition of the envelope \citep{kat97}.
Thus, we can follow the development of optical, UV, and 
supersoft X-ray light curves by a single modeled sequence
of steady wind solutions.

\subsection{System parameters of optically thick wind model}
     The light curves of our optically thick wind model are
parameterized by the WD mass ($M_{\rm WD}$), 
the chemical composition of the envelope, and the envelope mass 
($\Delta M_{\rm env, 0}$) at the outburst (day 0).
We have searched for the best fit model by changing these parameters,
for example, in a step of $0.05~M_\sun$ for the WD mass, 
of $0.01$ for hydrogen mass content, $X$, and of 0.05 for carbon,
nitrogen, and oxygen mass content, $C+N+O$.  It should be noted
here that hydrogen content $X$ and carbon, nitrogen, and oxygen
content $C+N+O$ are important because they are main players
in the CNO cycle but neon content is not because neon is not
involved in the CNO cycle.  The metal abundance of $Z=0.02$
is adopted, in which carbon, nitrogen, oxygen, and neon
are also included with the solar composition ratio.
We assume neon mass content of $Ne = 0.05$ \citep[taken from][]{van05},
because neon mass content cannot be determined only from 
our light curve fitting. 

\placefigure{v1974cyg_softXray}
\placefigure{v1974cyg_softXray_r_t}

\section{Light curve fitting}
\subsection{Supersoft X-ray and UV 1455 \AA~ fluxes}
\begin{figure}
\plotone{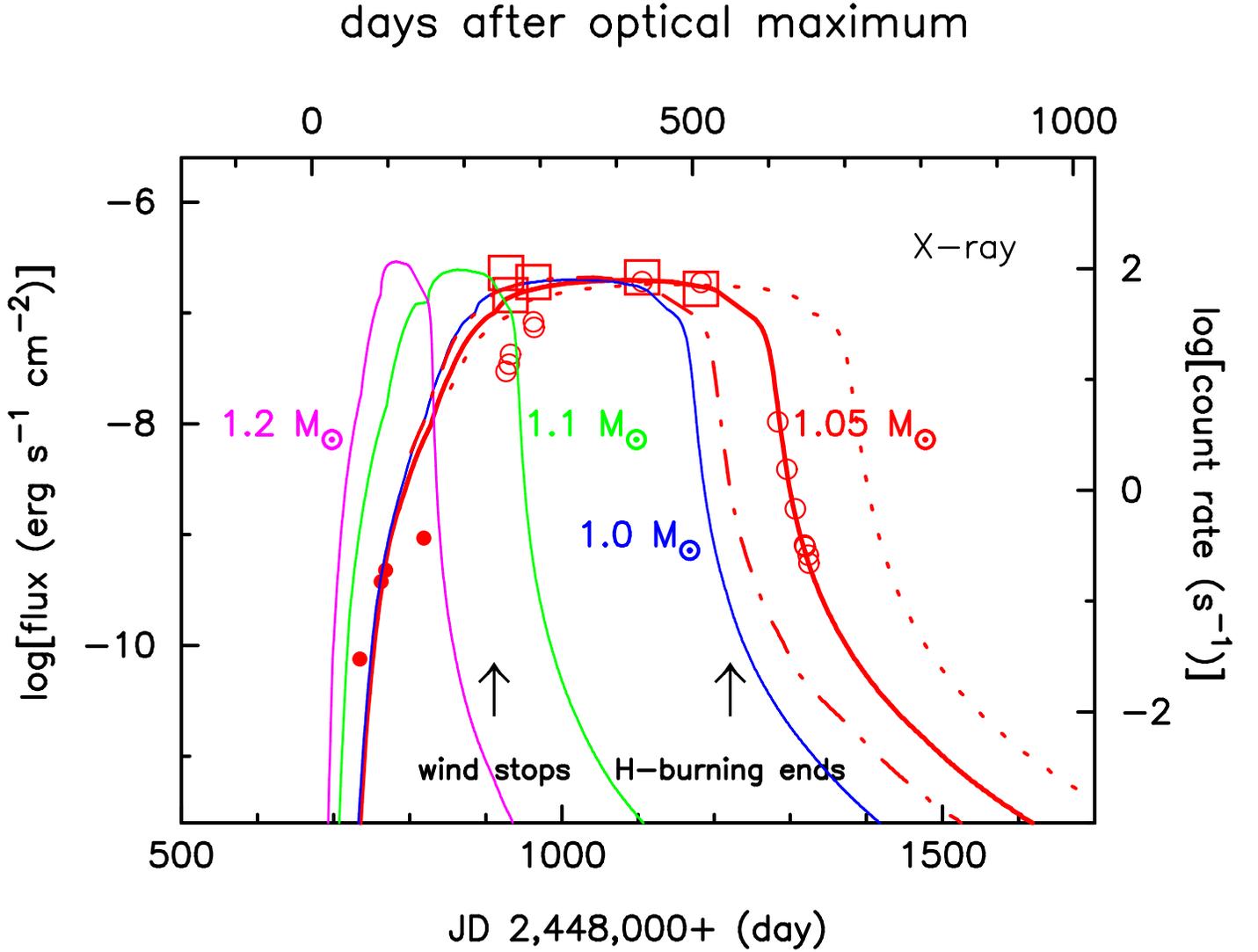}
\caption{
Calculated X-ray fluxes ($0.1-2.4$~keV) from white dwarf (WD)
photospheres are plotted against time for various WD masses and
chemical compositions together with {\it ROSAT} observation count rates 
\citep[{\it open and filled circles}: taken from][]{kra96}.
{\it Open circles}: dominated by soft X-rays.
{\it Filled circles}: dominated by hard X-rays.
{\it Open squares}: corrected X-ray fluxes \citep{bal98}.
The distance is assumed to be 2.2~kpc (see Discussion).
The epoch of the optical maximum corresponds to JD~2,448,673.67,
which is 8.67 days after the outburst.  
{\it Thin solid lines}: $1.0~M_\sun$,
$1.1~M_\sun$, and $1.2~M_\sun$ WDs with the envelope composition of
$X=0.35$, $C+N+O=0.30$, and $Ne=0.0$.
{\it Thick solid line}: the best-fit model of $1.05~M_\sun$ WD with
$X=0.46$, $C+N+O=0.15$, and $Ne = 0.05$.
Two other models are depicted for the same parameters as the best-fit
one except hydrogen content, $X=0.40$ ({\it dash-dotted line}) and
$X=0.53$ ({\it dotted line}).
Two epochs of the best-fit model are indicated by arrows.
\label{v1974cyg_softXray}}
\end{figure}

     {\it ROSAT} observation clearly shows that the supersoft
X-ray flux emerged on day $\sim 260$ after the outburst
and then decayed rapidly on day $\sim 600$ 
through a plateau phase of $\sim 300$ days \citep{kra96}.
Here, we define JD 2,448,665.0 as the outburst day,
8.67 days before the optical maximum (JD 2,488,673.67).
We have calculated many models, some of which are plotted 
in Figure \ref{v1974cyg_softXray} for the wavelength window
of $0.1 - 2.4$~keV.  Here, we have determined three parameters of
$M_{\rm WD}$, $X$, and $C+N+O$ by fitting three epochs with 
the observation, i.e., (1) when wind stops, (2) when hydrogen-burning
ends, and (3) when UV 1455 \AA~ flux reaches its maximum.  
We searched for the best fit model by eye.

     Our calculated X-ray fluxes in Figure \ref{v1974cyg_softXray}
show that the more massive the WD,
the shorter the duration of X-ray flat peak,
if the other two parameters are the same.
This is because a stronger gravity in more massive WDs
results in a smaller ignition mass.  As a result,
hydrogen is exhausted in a shorter period
\citep[see, e.g.,][for X-ray turn-off time]{kat97}.
On the other hand, if we increase hydrogen content,
we have a longer duration of hydrogen burning.
In this way, we choose the parameters that fit 
observed light curves.  The best fit model is $M_{\rm WD}= 1.05~M_\sun$,
$X=0.46$, $C+N+O= 0.15$, $Ne = 0.05$, 
and $\Delta M_{\rm env, 0} \approx 1.7 \times 10^{-5} M_\sun$, 
which is denoted by a thick solid line in Figure \ref{v1974cyg_softXray}.
Two epochs in the best-fit model are indicated by arrows:
(1) when the optically thick wind stops and 
(2) when the steady hydrogen-burning ends. 

Thin solid lines in Figure \ref{v1974cyg_softXray} depict
1.0, 1.1, and $1.2~M_\sun$ WDs with $X=0.35$, $C+N+O=0.30$, and
$Ne = 0.0$ while a thick solid line does the best-fit model 
of $1.05~M_\sun$ WD with $X=0.46$, $C+N+O=0.15$, and $Ne = 0.05$.
To see the effect of hydrogen content, we have added two other models 
with the same parameters as the best-fit
one except hydrogen content: $X=0.40$ ({\it dash-dotted line}) and
$X=0.53$ ({\it dotted line}).

     Figure \ref{v1974cyg_softXray} also shows that soft X-rays emerge
on day $\sim 250$.  In our model, soft X-rays appear after the wind
stops because the wind absorbs soft X-rays
\citep[e.g.,][]{sou96, hac03ka, hac03kb, hac03kc}.
The optically thick wind stops on day 245 just the time 
when the supersoft X-rays emerge.
After a plateau phase the X-ray flux quickly decreases 
as shown in Figure \ref{v1974cyg_softXray} because
the hydrogen shell-burning ends on day 558.

\begin{figure}
\plotone{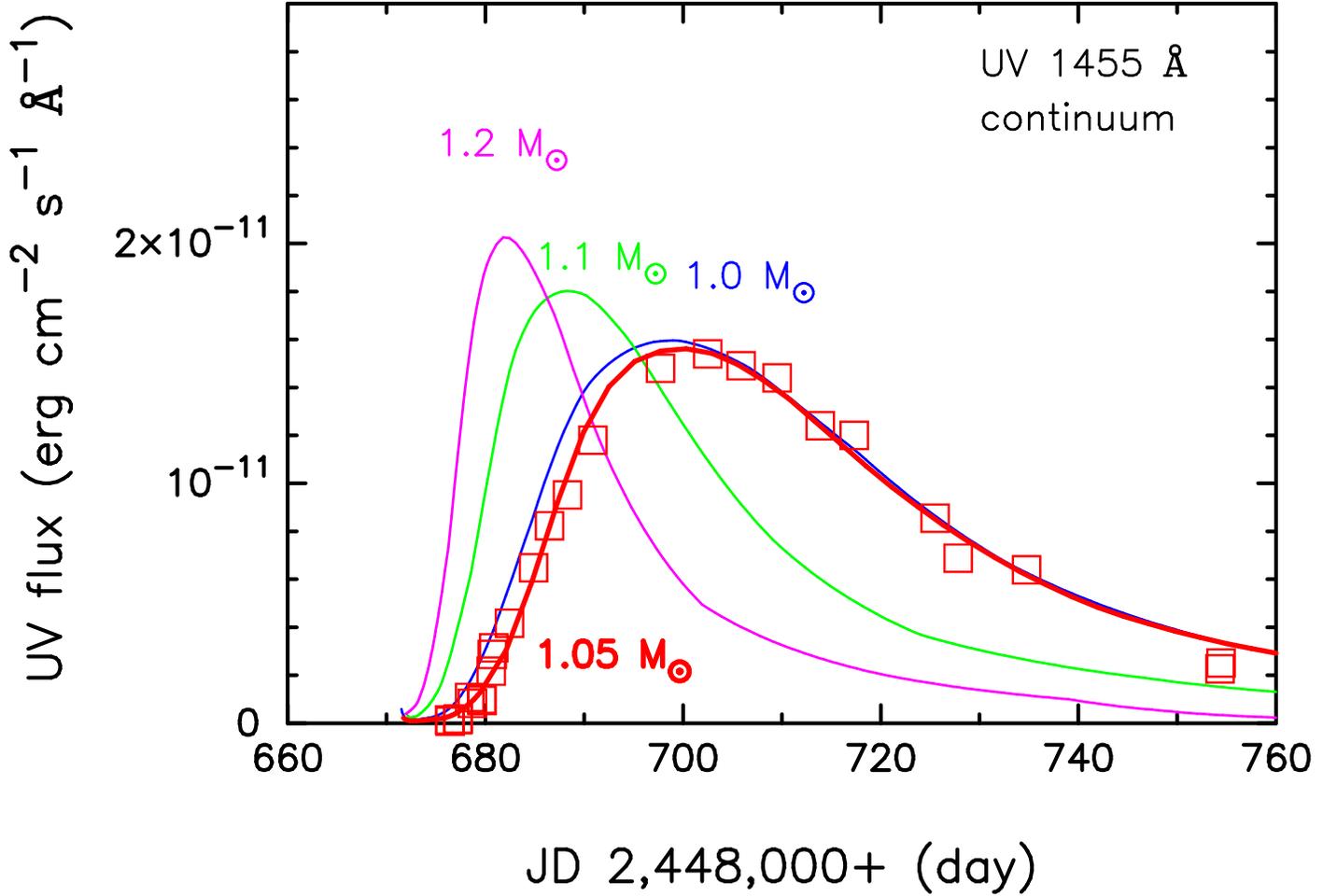}
\caption{
Calculated UV ($\lambda = 1455$~\AA~) fluxes are plotted 
together with the {\it IUE} observations 
\citep[{\it open square}: taken from][]{cas04}.
These curves correspond to the four models ({\it solid lines})
in Fig. \ref{v1974cyg_softXray}.
The distance of $d=1.7$~kpc is assumed.  Here we use
Seaton's (1979) absorption law of $A_\lambda = 8.3~E(B-V)= 2.65$
with an extinction of $E(B-V)= 0.32$ \citep{cho97}.
\label{v1974cyg_uv1455}}
\end{figure}

     The photospheric radius ($R_{\rm ph}$), temperature ($T_{\rm ph}$),
luminosity ($L_{\rm ph}$), and wind mass loss rate ({\it dashed line})
are plotted in Figure \ref{v1974cyg_softXray_r_t}.
Our results are roughly consistent with Balman et al.'s (1998) 
estimates for the photospheric radii and temperatures.
We should place the nova at a distance of 2.2~kpc to fit our
calculated X-ray flux with Balman et al.'s fluxes.
This distance is longer than that derived from the UV fitting
in Figure \ref{v1974cyg_uv1455}.  See discussion.

\placefigure{v1974cyg_uv1455}

     Figure \ref{v1974cyg_uv1455} depicts UV fluxes 
in a band of $\Delta \lambda=20$~\AA ~wide centered
at $\lambda=1455$~\AA , taken from \citet{cas04}.
The corresponding UV light curves are calculated 
for each model in Figure \ref{v1974cyg_softXray}.
Our best fitted $1.05~M_\sun$ WD model shows an excellent agreement
with the observation if we place the nova at a distance of 
1.7~kpc.  Here we adopt an absorption law given by \citet{sea79},
$A_\lambda= 8.3 E(B-V)= 2.65$,
together with an extinction of $E(B-V)= 0.32$ estimated by
\citet{cho97}.

\placefigure{tot_wind_ejected_visual}

\subsection{Optical fluxes}
     We cannot fit the observed visual light curve by our best fitted
model or even by other models with other sets of WD mass and envelope
chemical composition.  Therefore, we interpret that the optical flux
is dominated by free-free emission of the optically thin ejecta that
exist outside the photosphere.

\begin{figure}
\plotone{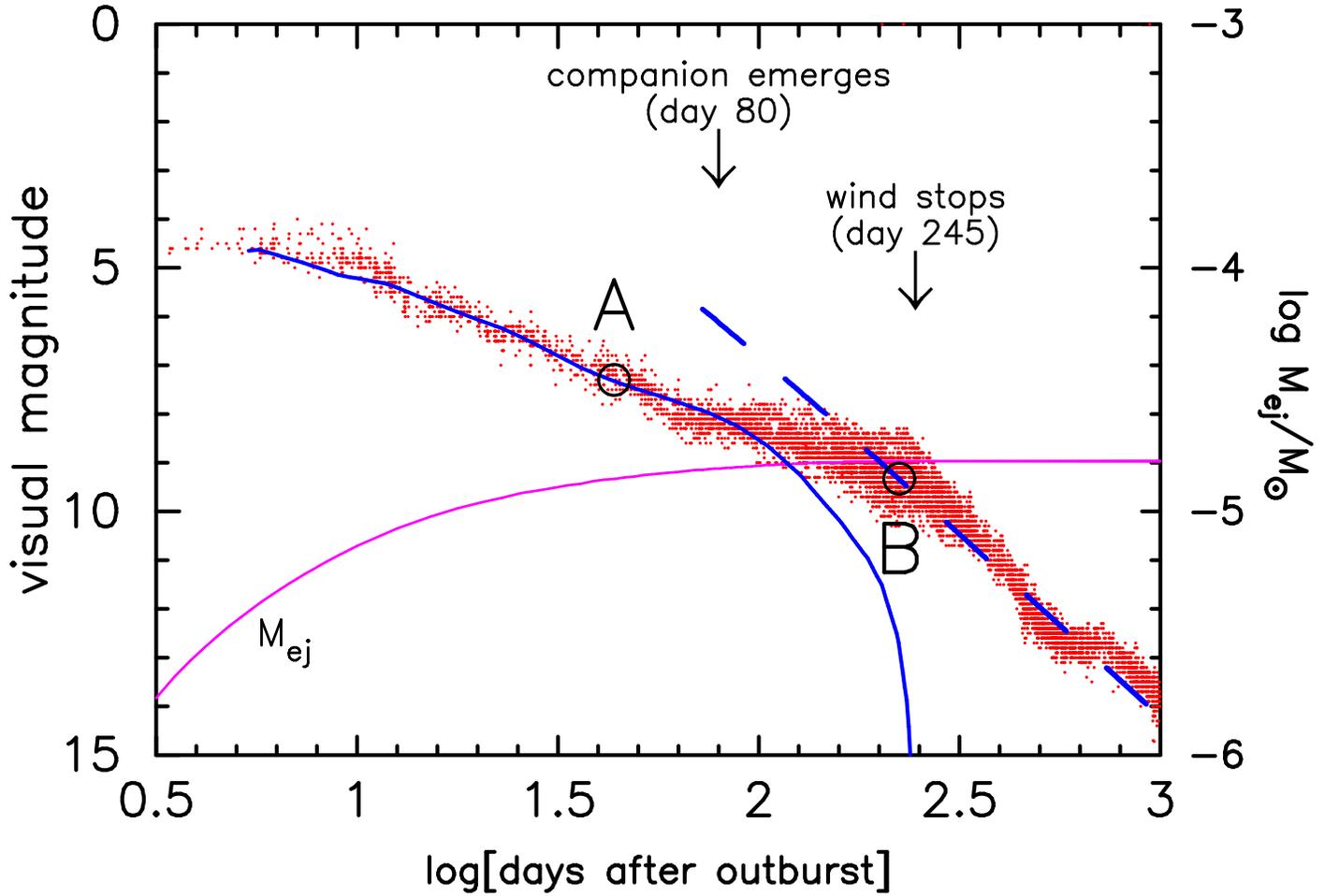}
\caption{
{\it Thick solid line:} visual magnitude of 
free-free emission from the optically thin ejecta, based on
equation (\ref{free-free-wind}): scaled to fit at day 43 (point A).
The flux decays with a slope of $\sim t^{-1.5}$ until day $\sim 100$.
{\it Dashed line:} free-free emission with a slope of $\sim t^{-3}$
after the wind stops: scaled to fit at day 224 (point B).
{\it Thin solid line:} ejected mass ($M_{\rm ej}$) from the WD by
the optically thick winds.  Here, we assume
JD~2,448,665.0 as the date of outburst.
{\it Small dots:} observational magnitudes taken from AAVSO archive.
Two epochs of the nova outburst are indicated by arrows:
the companion emerges from the WD photosphere and the
optically thick nova wind stops.
\label{tot_wind_ejected_visual}}
\end{figure}

     For the free-free emission of optically thin ejecta,
optical flux can be roughly estimated as
\begin{equation}
F_\lambda \propto \int N_e N_i d V 
\propto \int_{R_{\rm ph}}^\infty {\dot M_{\rm wind}^2 \over r^4} r^2 dr
\propto {\dot M_{\rm wind}^2 \over R_{\rm ph}}
\label{free-free-wind}
\end{equation}
during the optically thick wind phase,
where $F_\lambda$ is the flux at the
wavelength $\lambda$, $N_e$ and $N_i$ the number densities of 
electron and ion, $V$ the volume of the ejecta, 
$\dot M_{\rm wind}$ the wind massless rate.  Here, we use
the relation of $\rho_{\rm wind} = \dot M_{\rm wind}/ 4 \pi r^2 
v_{\rm wind}$, and $\rho_{\rm wind}$ and $v_{\rm wind}$ are 
the density and velocity of the wind, respectively.  
After the wind stops, we obtain
\begin{equation}
F_\lambda \propto \int N_e N_i d V 
\propto \rho^2 V \propto {M_{\rm ej}^2 \over V^2} V~
(\propto R^{-3} \propto t^{-3}),
\label{free-free-stop}
\end{equation}
\citep[e.g.,][]{woo97}, where $\rho$ is the density, $M_{\rm ej}$ 
the ejecta mass (in parenthesis, if $M_{\rm ej}$ is constant in time),
$R$ the radius of the ejecta ($V \propto R^3$), and $t$ the time after
the outburst.  Here, we substitute $\dot M_{\rm wind}$ and 
$R_{\rm ph}$ of our best fit model for those 
in equation(\ref{free-free-wind}).
We cannot uniquely specify the constant in equations 
(\ref{free-free-wind}) and (\ref{free-free-stop})
because radiative transfer is not calculated
outside the photosphere.  Instead,
we choose the constant to fit the light curve on day 43 denoted by A
(on the {\it thick solid line}) and on day 245 denoted 
by B (on the {\it dashed line})
in Figure \ref{tot_wind_ejected_visual}.  These two light curves
represent well the early/late parts of the observational data of AAVSO.

\citet{woo97} summarized the optical and infrared (IR) observations
of V1974 Cyg and concluded that 
$0.55 \mu$m $V$, $1.25 \mu$m $J$, $1.6 \mu$m $H$, and $2.3 \mu$m $K$
light curves all showed an abrupt transition from 
a $t^{-1.5}$ slope to a $t^{-3}$ slope at day $\sim 170$.
This $t^{-1.5}$ slope is very close to
the slope of our free-free light curve until day $\sim 100$.
After the wind stops, we have a slope of $t^{-3}$ 
as shown in Figure \ref{tot_wind_ejected_visual}.
This transition probably occurs when the optically thick
wind stops.  Therefore, our model is very consistent with
the temporal optical and IR observations.


\section{Discussion}
\subsection{Hard X-ray component}
     {\it ROSAT} observation shows that hard X-ray flux
increases on day $70-100$ and then decays on day $270-300$.
This hard component is suggested to be shock-origin
between ejecta \citep{kra96}.  Here we present another idea
that these hard X-rays are originated from a shock
between the optically thick wind and the companion as described below.

\begin{figure}
\plotone{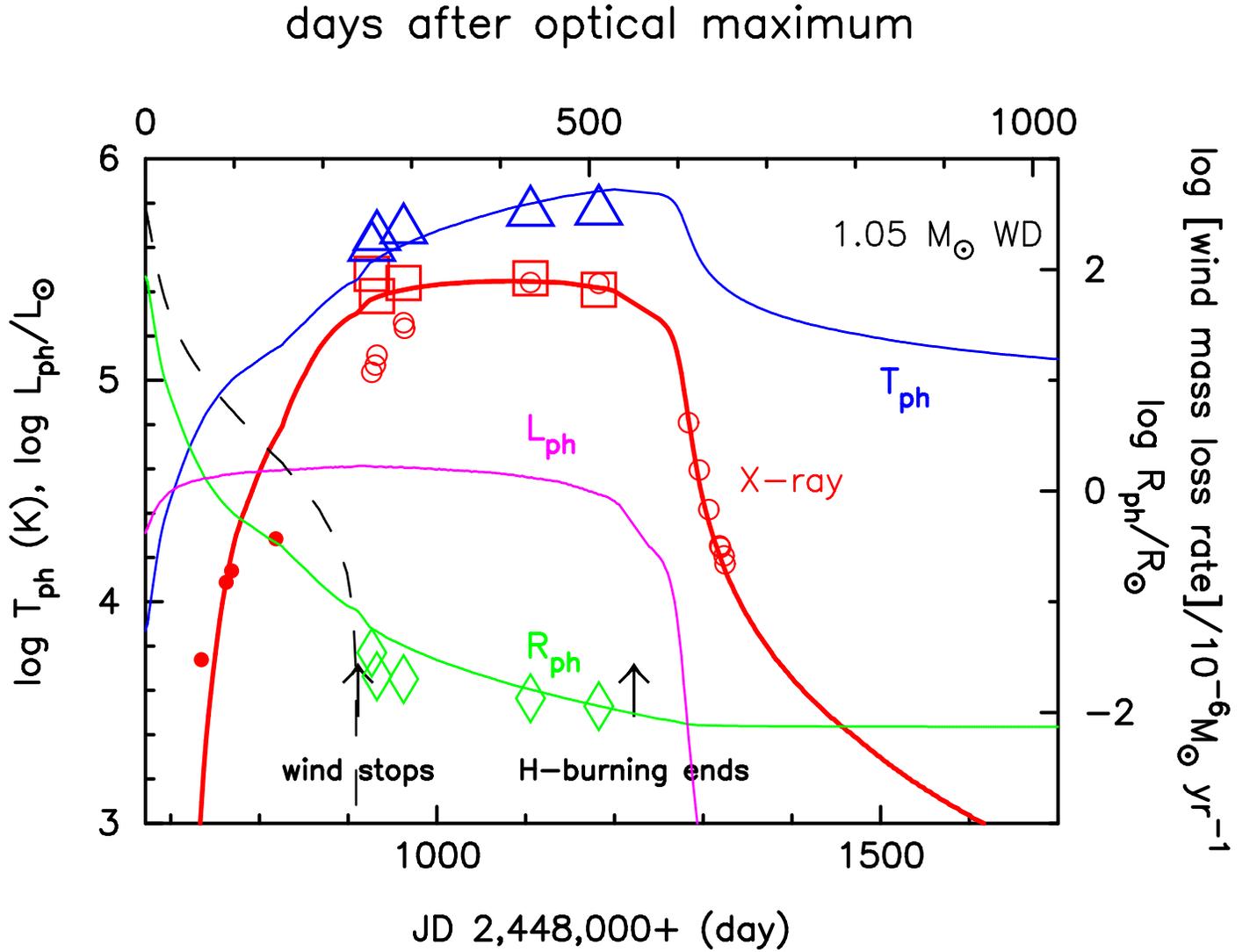}
\caption{
Wind mass loss rate ({\it dashed line}),
photospheric temperature ($T_{\rm ph}$), 
photospheric radius ($R_{\rm ph}$), 
photospheric luminosity ($L_{\rm ph}$), and
X-ray flux ({\it thick solid line}) of
the best fit model ($M_{\rm WD}= 1.05~M_\sun$,
$X=0.46$, $C+N+O =0.15$, and $Ne =0.05$).
X-ray fluxes ({\it Open squares}), photospheric temperatures 
({\it Open triangles}), and photospheric radii ({\it Open diamonds}).
These are taken from \citet{bal98}.  Here, photospheric radii 
are calculated from $A_1 \equiv (R_{\rm ph} / d )^2$
in Balman et al.'s Table 1 with the distance of $d = 1.7$~kpc.
{\it Open and filled circles}: the same X-ray count rates as
in Fig. \ref{v1974cyg_softXray}.  See discussion for more details.
\label{v1974cyg_softXray_r_t}}
\end{figure}

     V1974 Cyg is a binary system with an orbital period of 
$P_{\rm orb} = 0.0812585$~days \citep[e.g.,][]{dey94, ret97}.
\citet{par95} and \citet{ret97} estimated the companion mass
at $0.21 ~M_\sun$ from this orbital period.
Using these values we obtain the separation,
$a= 0.853 ~R_\sun$, the effective radii of each Roche lobe,
$R_1^* = 0.444 ~R_\sun$ and $R_2^* = 0.215 ~R_\sun$ for
the primary (WD) and the secondary component, respectively.
Our optically thick wind model predicts that the companion star
emerges from the WD photosphere about day 80 
(for the photospheric radius, see Fig. \ref{v1974cyg_softXray_r_t}).

Before day $\sim 80$, the companion resides deep inside 
the WD photosphere and we do not detect hard X-rays.
After the companion emerges from the WD photosphere,
the shock front can be directly observed.
The optically thick wind stops on day 245
and we expect that the hard X-ray component decays after that.
This hard X-ray flux may show orbital modulations
if the inclination angle of binary is large enough.  However,
\citet{cho97} estimated it at $i \sim 39 \arcdeg$.
For such a small inclination angle, 
we are able to see main parts of the shock front at any binary phase,
because a bow-shock is formed off the surface of the companion 
\citep[see, e.g.,][]{shi86} and basically optically thin to hard X-rays.
Therefore, orbital modulation of hard X-ray flux is hardly observed,
which is consistent with the observation \citep{kra96}.

\citet{bal98} estimated the hydrogen column density of 
the hard X-ray component and concluded that it decreases,
by a factor of $\sim 10$,
from $N_{\rm H} \sim 10^{22.2}$ to $10^{21.3}$~cm$^{-2}$ between
day 70 and day 260, and almost constant after that.
In our optically thick wind model, the neutral hydrogen column
density is given by
\begin{equation}
N_{\rm H} \propto {{X} \over {m_{\rm H}}} \int_{r_{\rm s}}^\infty 
\rho_{\rm wind} d r
\approx {{\dot M_{\rm wind} X} \over {4 \pi a v_{\rm wind} m_{\rm H}}}
\propto \dot M_{\rm wind},
\end{equation}
where $m_{\rm H}$ is the
mass of hydrogen atom, $r_{\rm s}$ the position of the bow-shock from
the WD center, $a$ the separation of the binary, and
we roughly assume that the bow-shock front is at distance of
$r_{\rm s} = (0.5 - 0.7)a$
from the WD center.  Our wind mass loss rate
decreases from $\sim 10^{-5}$ to $\sim 10^{-6} M_\sun$~yr$^{-1}$
between day 70 and day 260 (see Fig. \ref{v1974cyg_softXray_r_t}),
which is very consistent with Balman et al.'s results.

\subsection{WD mass and chemical composition}
     Several groups estimated the WD mass of V1974 Cyg.
\citet{ret97} gave a mass of $M_{\rm WD} = 0.75 - 1.07 ~M_\sun$
based on the precessing disk model of superhump phenomenon.
A similar range of $0.75 - 1.1 ~M_\sun$ is also obtained by
\citet{par95} from various empirical relations on novae.
Very recently, \citet{sal05} found the WD mass to be $0.9 ~M_\sun$ for
50\% mixing of a solar composition envelope with a O-Ne degenerate
core, or $1.0 ~M_\sun$ for 25\% mixing,
by comparing the evolutional speed of post-wind phase of V1974 Cyg
with their post-wind phase of static envelope solutions.
Their values for the $1.0 ~M_\sun$ are roughly consistent with our results. 

     \citet{van05} criticized \citet{aus96} results and reanalyzed
chemical abundances of the ejecta from optical and UV spectra.
They obtained that He$= 1.2 \pm 0.2$, C$= 0.7 \pm 0.2$, N$= 44.9 \pm 11$,
O$= 12.8 \pm 7$, and Ne$= 41.5 \pm 17$ by number relative
to hydrogen and relative to solar.  In our notation,
these correspond to $X= 0.55$, $Y= 0.25$, $C+N+O= 0.12$,
$Ne = 0.06$, and $Z=0.02$ by mass weight.
The hydrogen content is a bit higher
but these values are very consistent with our results.

\subsection{Distance}
     We estimate the distance to V1974 Cyg 
from the UV 1455 \AA ~light curve fitting.
The absorption at $\lambda = 1455$ \AA ~ is calculated to be
$A_\lambda = 8.3 ~E(B-V) = 2.65$ \citep[e.g.,][]{sea79},
where we adopt the absorption at the visual band, 
$A_V = 3.1 ~E(B-V)= 0.99$ \citep{cho97}.  
Then we have a distance to the nova of $d \approx 1.7$~kpc.
For the X-ray band, \citet{bal98} obtained $(R_{\rm ph} / d )^2 =
(0.22-0.26) \times 10^{-25}$ on day 518 (corresponding to their day 511).
Using $R_{\rm ph} = 0.0115~R_\sun$ on day 518 of our best fit model,
we obtain the distance of $d = 1.6 - 1.7$~kpc, which is consistent 
with our distance estimation from the UV fitting.
On the other hand, our X-ray flux combined with Balman et al.'s (1998)
fluxes gives a rather large distance of 2.2~kpc.
This difference may come from the different model
parameters adopted in their atmosphere models: $1.2~M_\sun$ WD 
and $X= 0.54$, $Y= 0.21$, $Z= 0.02$, $C= 0.002$, $O= 0.103$, 
$N= 0.002$, and $Ne= 0.123$.  Their neon mass is much higher 
than the observation $Ne=0.06$ \citep{van05}.
Therefore we take the distance of $d = 1.7$~kpc.
These distances are all within the range listed in \citet{cho97},
$d= 1.3 - 3.5$~kpc with a most probable value of 1.8~kpc,
derived mainly from maximum magnitude-rate of decline (MMRD) relations.




\acknowledgments
     We thank A. Cassatella for providing us with their 
machine readable UV 1455 \AA ~data of V1974 Cygni and also 
AAVSO for the visual data of V1974 Cygni.  We are also grateful
to the anonymous referee for useful comments to improve the manuscript.
This research has been supported in part by the Grant-in-Aid for
Scientific Research (16540211, 16540219) 
of the Japan Society for the Promotion of Science.

\clearpage

\clearpage

\clearpage

\clearpage


\end{document}